\documentclass[12pt]{article}
\usepackage{graphics}
\usepackage[square,sort,comma,numbers,compress]{natbib}
\usepackage[toc,page]{appendix}
\usepackage{graphicx}
\usepackage{subfig}
\usepackage{hyperref}
\usepackage{array}
\usepackage{amsmath}
\usepackage{amssymb}
\usepackage{comment}
\usepackage{float}
\usepackage{graphicx} %Include figure filesusepackage{graphicx} %Include figure files
\usepackage{feynmf}
\usepackage[compat=1.1.0]{tikz-feynman}
\usepackage{tablefootnote}
\usepackage{authblk}

\usepackage{multirow}
\usepackage{hhline}
\usepackage{xcolor}

\newcommand{\ZN}{\mathbb{Z}_N}
\numberwithin{equation}{section}

\newcommand\eea{\end{eqnarray}}
\newcommand\bea{\begin{eqnarray}}

\begin{document}

%\date{Today}
%%%%%%%%%%%%%%%%%%%%
\title{{\bf{\Large Non-linearly realized discrete symmetries }}}
%%%%%%%%%%%%%%%%%%%%

\author[1]{\normalsize Saurav Das\thanks{sauutsab@umd.edu}}
\author[1]{\normalsize Anson Hook \thanks{hook@umd.edu}}
\affil[1]{Maryland Center for Fundamental Physics, University of Maryland, College Park, MD 20742, USA}

\date{}

\maketitle
\begin{abstract}
\noindent 
While non-linear realizations of continuous symmetries feature derivative interactions and have no potential, non-linear realizations of discrete symmetries feature non-derivative interactions and have a highly suppressed potential.  These Goldstone bosons of discrete symmetries have a non-zero potential, but the potential generated from quantum corrections is inherently very highly suppressed.  We explore various discrete symmetries and to what extent the potential is suppressed for each of them.

\end{abstract}
\vskip 1cm

%%%%%%%%%%%%%%%%%%%%%%%%%%%%%%%%% Introduction %%%%%%%%%%%%%%%%%%%%%%%%%%%%%%%%%%%%%%%%%%%%%%%%%%%%
%\begin{multicols}{2}
\section{Introduction}

%%%%%%%%%%%%%%%%%%%%%%%%%%%%%%%%%%%%%%%%%%%%%%%%%%%%%%%%%%%%%%%%%%%%%%%%%%%%%%%%%%%%%%%%%%%

Non-linear realizations of continuous symmetries, Nambu Goldstone bosons (NGBs), often appear in particle physics models.  
Perhaps the most famous example of a pseudo-Nambu Goldstone boson (pNGB) appears in the Standard Model and is the pion.  
Exact Goldstone bosons are highly constrained by their
continuous shift symmetry so that they are derivatively coupled and do not have a potential.  Goldstone bosons are interesting objects and there exists a vast literature studying them~\cite{Nambu:1961tp, Goldstone:1961eq,  Goldstone:1962es, Adler:1964um, Nielsen:1975hm, Brauner:2010wm, Watanabe:2012hr, Cheung:2014dqa, Low:2014nga, Cheung:2016drk, Low:2017mlh, Low:2018acv, Rodina:2018pcb}.

While interesting in their own right, the symmetries of a Goldstone boson are often too restrictive to be useful and they are often made into pNGBs by explicitly breaking their exact shift symmetry.  
The breaking of the exact shift symmetry can reintroduce unwanted features such as a large mass term, typically referred to as the Hierarchy Problem.  Unsurprisingly, many of the features discussed below will have analogues with various solutions to the Hierarchy problem that we will only briefly touch upon, as there exists a whole class of models where the Higgs is a pNGB~\cite{ArkaniHamed:2001nc, ArkaniHamed:2002qx, ArkaniHamed:2002qy, Contino:2003ve, Agashe:2004rs}.

In addition to solutions to Hierarchy problems, discrete symmetries are also ubiquitous in models of flavor, see e.g. Refs.~\cite{Ishimori:2010au, Altarelli:2010gt} and references therein.  These flavor models use group theoretic properties to explain various observed properties of the quark and lepton mass matrices.  These models typically involve spontaneous breaking of the discrete symmetries, leading to their non-linear realizations.  The results discussed in this paper will help explain features such as anomalously light scalars that appear in these models.

%The lack of a potential is an appealing starting point when building models where terms in the potential, such as the mass term, are required to be small.  By starting with a Goldstone boson, one can break the continuous shift symmetry and generate a mass in a controlled manner.

In this article, we initiate a study of the non-linear realizations of discrete symmetries, objects we dub ``discrete" NGBs.
We find that discrete NGBs combine the best features of both exact Goldstone bosons and ordinary NGBs.
On one hand, discrete NGBs can have large $\mathcal{O}(1)$ Yukawa couplings.
On the other hand, the potential radiatively generated from these Yukawa couplings is typically very highly suppressed.
%While having a feature such as this is not too surprising, after all in the limit where the discrete shift symmetry becomes continuous the potential must vanish, what is surprising is that the rate at which the potential vanishes tends to be expoen
%e.g. a highly suppressed potential, while allowing one to write down large Yukawa couplings that while do not respect a continuous shift symmetry, do respect a discrete shift symmetry.
These features can all be understood in the simplest example of a non-linear realization of the abelian discrete symmetry $\ZN$, which we now briefly review~\cite{Hook:2018jle}.

A non-linearly realized $\ZN$ features a periodic scalar $\pi_0$ with a period $2 \pi f$ so that $\pi_0 = \pi_0 + 2 \pi f$.  Under the $\ZN$ symmetry, $\pi_0$ transforms as
\bea
\frac{\pi_0}{f} \rightarrow \frac{\pi_0}{f} + \frac{2 \pi}{N}. \nonumber
\eea
Like Goldstone bosons and pNGBs, it is useful to exponentiate the scalar to obtain a field that transforms linearly under the $\ZN$ symmetry.  In this case, we introduce the field $\phi = f e^{i \pi_0/f}$ that transforms as $\phi \rightarrow e^{2 \pi i/N} \phi$ under the $\ZN$ symmetry.
To see that $\pi_0$ should have a suppressed potential, we simply need write down the leading order term in its potential.  Most of the first terms one can write, preserve an accidental $U(1)$ symmetry and do not give $\pi_0$ a mass.  The leading order analytic piece that can give a mass to $\pi_0$ can easily be seen to be
\bea
V \sim \phi^N \nonumber
\eea 
To generate a potential, one must first generate the operator $\phi^N$.  If the Yukawa coupling appears as $y \phi$, then each $\phi$ is accompanied by a Yukawa coupling so that the mass term generated must scale as $y^N$. Thus, despite a potentially large Yukawa coupling, the potential is exponentially suppressed.

To see this explicitly, we couple a set of N fermions to $\pi_0$ in a $\ZN$ symmetric manner.  We introduce N fermions $\psi_1 \cdots \psi_N$ 
that are exchanged cyclically under the $\ZN$ symmetry, $\psi_1 \rightarrow \psi_2 \rightarrow \psi_3 \cdots \psi_N \rightarrow \psi_1$. The leading order yukawa coupling that can be written is
\bea
\sum_{j=1}^N \left ( m_\psi+ \frac{y}{2}  e^{i (\frac{2 \pi j}{N} - \frac{\pi}{2} )} \phi + \frac{y}{2}  e^{-i (\frac{2 \pi j}{N} - \frac{\pi}{2} )} \phi^\dagger \right ) \psi_j \psi^c_j =  \sum_{j=1}^N \left ( m_\psi + y f \sin \left ( \frac{\pi_0}{f} + \frac{2 \pi j}{N} \right ) \right ) \psi_j \psi^c_j, \nonumber
\eea
where we have taken the Yukawa coupling $y$ to be real.  From this, one can calculate the one-loop Colemann Weinberg potential of $\pi_0$ and find that the leading order contribution scales as
\bea
V(\phi) \sim m_\psi^4  \left ( \frac{y f}{m_\psi} \right )^N \cos \left(\frac{N \pi_0}{f}\right) . \nonumber
\eea
As expected from the general arguments given before, we see that the potential is suppressed by $y^N$.  The proper expansion parameter is $(y f/m_\psi)$ as opposed to $y$ as one is simply doing a Taylor series of the fermion mass
\bea
m_{\psi,j} (\pi_0) = m_\psi \left ( 1 + \frac{y f}{m_\psi} \sin \left ( \frac{\pi_0}{f} + \frac{2\pi j}{N}  \right ) \right ). \nonumber
\eea
Thus we see that if the expansion parameter is small, then as N increases, $\ZN$ becomes an exponentially good approximation to a $U(1)$ and the corresponding discrete NGB mass goes to zero exponentially quickly~\footnote{Depending on the details of the theory, the $y^N$ can also be understood as collective symmetry breaking~\cite{ArkaniHamed:2001nc, ArkaniHamed:2002qy}.}.
From this simple example, it can be seen that non-linear realizations of discrete symmetries can feature amazing cancellations that result in highly suppressed potentials.

In this article, we will study non-linear realizations of non-abelian discrete symmetries and to what extent their discrete NGBs have their potentials suppressed.
Discrete NGBs of non-abelian discrete symmetries have many of the same features of as their abelian cousins.  As with the abelian case, the crucial point in determining how suppressed the potential is, is to determine the dimension of the operator which gives the discrete NGBs a mass.

When considering continuous non-abelian Lie groups, one must specify the breaking pattern in order to determine the number of Goldstone bosons or equivalently one must specify the representation doing the symmetry breaking.
Analogously, when dealing with non-abelian discrete NGBs, one must also specify the representation doing the breaking.  A surprising feature of non-abelian discrete groups is that they can approximate many different groups and cosets to varying degrees of accuracy.
%The most surprising novelty with non-abelian discrete Goldstone bosons is that fixing the group does not determine the number of discrete Goldstone bosons.  There are many ways in which to non-linearly realize a discrete non-abelian Goldston boson.
To see this feature in action, assume that you have a scalar $\phi$ in an M dimensional real representation of a non-abelian discrete symmetry $\mathcal{G}$ and take the potential for $\phi$ to include a negative mass squared term.  The largest accidental continuous symmetry that can act on this M dimensional representation is an $SO(M)$ symmetry and we take the leading order operator that breaks this accidental $SO(M)$ symmetry to be $\phi^{N_M}$.
Thus there are $M-1$ discrete NGBs which non-linearly realize $\mathcal{G}$ and approximate the continuous coset $SO(M)/SO(M-1)$.
This situation is in complete analogy to $Z_N$ where we took a 2 dimensional real representation that had a $SO(2)$ accidental symmetry.  The leading order operator that breaks this accidental $SO(2)$ symmetry was $\phi^{N}$.

If the discrete NGBs obtain an $SO(M)$ breaking but $\mathcal{G}$ preserving Yukawa coupling via the interaction $y \phi \overline \Psi \Psi$, then by the same arguments used before in the $\ZN$ example, the potential giving a mass term to the $M-1$ discrete NGBs scales as $y^{N_M} \phi^{N_M}$.
The larger $N_M$ is, the better the discrete NGBs approximates the real Goldstone bosons of $SO(M)/SO(M-1)$.
If $N_M > 2$, it is a good enough approximation to remove the quadratic divergence.  Because there are many different representations with many different dimensions that all approximate different continuous groups, we see that non-abelian discrete groups can approximate as many continous cosets as they have representations~\footnote{Interestingly by exactly the same reasoning, large dimensional representations of continuous symmetries can also be used to approximate various cosets.}.  This scenario is in complete analogy to Twin Higgs models~\cite{Chacko:2005pe} or Twin Higgs-like models~\cite{Craig:2014aea} where a $\mathcal{Z}_2$ symmetry plus gauge invariance forces the Higgs mass term to be accidentally $SO(4)$ symmetric and a mass for the pNGB Higgs is only generated by a $SO(4)$ breaking quartic term.

In Sec.~\ref{Sec: explicit example}, we give a simple $A_4$ example and work out in detail how the cancellations occur.
In Sec.~\ref{Sec: invariance}, we describe how the results of invariant theory can be used to obtain how suppressed a potential is for a generic non-abelian discrete symmetry.
In Sec.~\ref{Sec: exchange}, we explore how exchange representations of discrete symmetry groups can be used.
Finally, we conclude in Sec.~\ref{Sec: conclusion}.

%%%%%%%%%%%%%%%%%%%%%%%%%%%%%  Section 2     %%%%%%%%%%%%%%%%%%%%%%%%%%%%%%%%%%%%%%%%%%%%%%
\section{Explicit example} \label{Sec: explicit example}

As a simple example, we will first consider the case of an $A_4$ non-abelian discrete symmetry.
Consider a scalar $\phi$ which transforms as a triplet under $A_4$ \cite{Ishimori:2010au}.  $A_4$ is the group of all even permutations of four objects and is isomorphic to proper rotations of a regular tetrahedron.  From this, one can geometrically see that it is a finite subgroup of $SO(3)$. The scalar $\phi$ is coupled to a Dirac fermion $\Psi$ via a Yukawa coupling.  For simplicity, $\Psi$ is also taken to be a triplet.  As we are focusing on the quantum generated potential, we will take the tree level potential to be $SO(3)$ symmetric except for the Yukawa interaction~\footnote{If this assumption bothers the reader, one can start with a more complicated example where the renormalizable potential automatically preserves an accidental global symmetry, e.g. the doublet representation of $T'$.  Alternatively, one can simply assume that the UV theory gives an approximate $SO(3)$ symmetry at tree level in analogy with chiral perturbation theory.}.
\begin{equation}
\begin{split}
&\mathcal{L}_{\text{tree}} =\mathcal{L}_{\text{kin}}+\mathcal{L}_{\text{V}(\phi)} +\mathcal{L}_{\text{int}}\\
&\mathcal{L}_{\text{kin}}=\frac{1}{2}\partial^{\mu}\phi^T\partial_{\mu}\phi+\overline{\Psi}(i\gamma^{\mu}\partial_{\mu})\Psi\\
&\mathcal{L}_{\text{V}(\phi)}=\frac{m^2}{2}\phi^T\phi-\frac{\lambda}{4}(\phi^T\phi)^2 \\
\end{split}
\end{equation}
The tachyonic mass generates a tree level set of degenerate vacua with $\langle\phi^T\phi\rangle=\frac{m^2}{\lambda}\equiv f^2 $ that is spanned by the usual Goldstone bosons. After spontaneous symmetry breaking, the scalar triplet around the vacuum $(0,0,f)$ is parameterized by
\begin{equation}
\phi=\exp\Bigg[\frac{1}{f}{\begin{pmatrix}
0&0&\pi_1\\
0&0&\pi_2\\
-\pi_1&-\pi_2&0\\
\end{pmatrix}}\Bigg]\begin{pmatrix}0\\0\\f\end{pmatrix} ,
\end{equation}
where $\pi$ are the familiar Goldstone bosons of the breaking $SO(3)/SO(2)$ and will later become discrete NGBs of $A_4$.
Under the tree level $SO(3)$ symmetry, the pions have a shift symmetry which forbids non-derivative couplings. 
The approximate $SO(3)$ symmetry is broken explicitly by the Yukawa coupling so that the radiative corrections only respect the global $A_4$ symmetry instead of the larger $SO(3)$ symmetry and generate a mass for the pions.

The pions $\pi_1$ and $\pi_2$ provide a non-linear realization of the $A_4$ symmetry.  $A_4$ has two generators, $s$ and $t$, with $s^2=t^3=(st)^3=e$.  Explicitly, the $s$ and $t$ generators in the triplet representation are
\begin{equation}
\begin{split}
    s=&\begin{pmatrix}
    1&0&0\\
    0&-1&0\\
    0&0&-1\\
    \end{pmatrix}\\
    t=&\begin{pmatrix}
    0&0&1\\
    1&0&0\\
    0&1&0\\
    \end{pmatrix} .
\end{split}
\end{equation}
To leading order in the pions, the $s$ generator is realized by sending $\pi_2 \rightarrow \pi_2 \pm \pi f$. Again to leading order in the pions, the $t$ generator is realized by sending $\pi_1 \rightarrow \pi_1 + \pi f/2$ followed by $\pi_2 \rightarrow \pi_2 - \pi f/2$. 
%\AH{Added this short paragraph.}The pions $\pi_1$ and $\pi_2$ provide a non-linear realization of the $A_4$ symmetry.  $A_4$ has two generators S and T.  The T generator is realized by sending $\pi_1 \rightarrow \pi_1 + \pi f/2$ followed by $\pi_2 \rightarrow \pi_2 - \pi f/2$.  The S generator is realized by sending $\pi_1 \rightarrow \pi_1 + \pi f$ followed by $\pi_2 \rightarrow \pi_2 + \pi f$.

\begin{figure}[H]
    \centering
    \includegraphics[width=\textwidth]{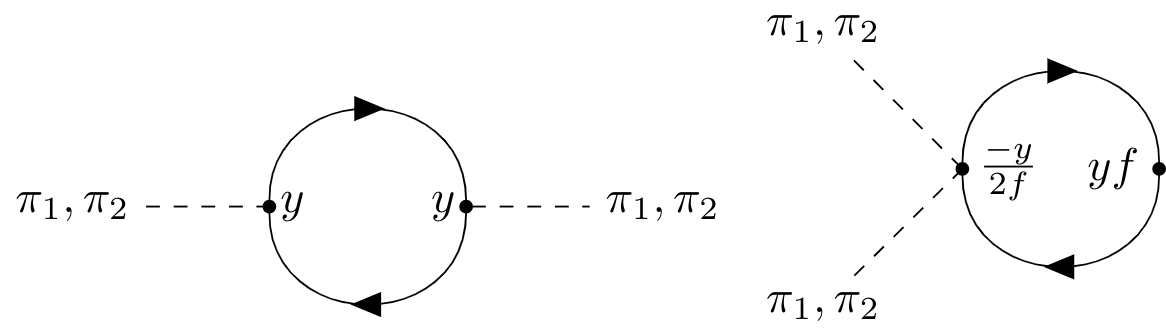}
    \caption{The two quadratically divergent one loop diagrams for pion mass that cancel each other exactly. }
    \label{fig:loop}
\end{figure}

The most general $A_4$ invariant Yukawa interaction can be written as
\begin{equation}
\begin{split}
&\mathcal{L}_{\text{int}}=\Bigg[y_s \begin{pmatrix} \{ \overline{\Psi}_2\Psi_3 \}\\ \{\overline{\Psi}_3\Psi_1\}\\ \{\overline{\Psi}_1\Psi_2\} \\ \end{pmatrix} + y_a \begin{pmatrix}   [\overline{\Psi}_2\Psi_3] \\ [\overline{\Psi}_3\Psi_1] \\ [\overline{\Psi}_1\Psi_2]  \\ \end{pmatrix}\Bigg]\cdot \phi
\end{split}
\end{equation}
where 
\begin{equation}
\begin{split}
\{A_i B_j\}=A_i B_j+B_j A_i \\
[A_i B_j]=A_i B_j-B_j A_i
\end{split}
\end{equation}
The anti-symmetric coupling, $y_a$, is the usual $SO(3)$ invariant piece, does not give a mass to the pions and will thus be neglected for the rest of the section.
The novel symmetric coupling, $y_s$, gives the pions a non-zero yukawa coupling to the fermions but at the same time protects it from the standard quadratic divergences. To see that explicitly, let us Taylor expand the scalar in terms of the pions.
\begin{equation}
\begin{split}
\phi_1 = \pi_1 \ \ \phi_2 = \pi_2 \ \ \phi_3 = f\bigg(1-\frac{1}{2} \frac{\pi_1^2+\pi_2^2}{f^2}\bigg)
\end{split}
\end{equation}
At this order, the interaction term becomes
\begin{equation}
\begin{split}
\mathcal{L}_{\text{int}}&=y\pi_1\big(\overline{\Psi}_2\Psi_3+\overline{\Psi}_3\Psi_2)+y\pi_2\big(\overline{\Psi}_3\Psi_1+\overline{\Psi}_1\Psi_3)\\
&+yf\bigg(1-\frac{1}{2} \frac{\pi_1^2+\pi_2^2}{f^2}\bigg)\big(\overline{\Psi}_1\Psi_2+\overline{\Psi}_2\Psi_1)
\end{split}
\end{equation}
where we have abbreviated the Yukawa coupling $y_s=y$.  One can calculate the one loop quadratic divergence from the two diagrams shown in Fig.~\ref{fig:loop}, which neatly cancel each other in a manner very reminisent of Little Higgs models~\cite{ArkaniHamed:2002qy,Perelstein:2003wd,Perelstein:2005ka,Schmaltz:2005ky}.

The surprising cancellation found in the previous example follows from symmetry. Let us write the interaction as
\begin{equation}
\begin{split}
L_{\text{int}}=M^{IJ}\overline{\Psi}_I\Psi_J\\
M=y\begin{pmatrix}
0&\phi_3&\phi_2\\
\phi_3&0&\phi_1\\
\phi_2&\phi_1&0\\
\end{pmatrix}
\end{split}
\end{equation}
where the position of the indices on the `fermion mass matrix' has no special meaning, $M_{IJ}=M^{IJ}$.
The quadraticly divergent part of the one loop Coleman-Weinberg potential for the scalar is
\begin{equation}
\begin{split}
V_{\text{1 loop}}&\supset -\frac{1}{4\pi^2}  \Lambda^2\; \text{Tr}[M\cdot M^T]\\
&=-\frac{1}{4\pi^2} y^2 \Lambda^2 \;2(\phi^T \phi)
\end{split}
\end{equation}
which protects the shift symmetry of the pions. This form is required as the only quadratic invariant of the $A_4$ group is also an $SO(3)$ invariant.
As this conclusion is simply group theoretic, it is not surprising that adding a vector like mass for the fermions does not change anything.  In presence of a vector like mass $m_\psi$, the quadratic part of the one loop potential is 
\begin{equation}
\begin{split}
V_{\text{1 loop}}&\supset-\frac{1}{4\pi^2}\Lambda^2 \text{Tr}[M\cdot M^T]\\
&=-\frac{1}{4\pi^2}\Lambda^2\left[y^2\; 2(\phi^T \phi)+3m_\psi^2\right]\\
M&=y\begin{pmatrix}
0&\phi_3&\phi_2\\
\phi_3&0&\phi_1\\
\phi_2&\phi_1&0\\
\end{pmatrix}+m_\psi I_{3\times 3}
\end{split}
\end{equation}
which also doesn't introduce a potential for the pions. Since the Dirac mass term trivially respects the $SO(3)$ symmetry, the reader may have anticipated this behavior.

\begin{figure}[t]
    \centering
    \includegraphics[width=0.45\textwidth]{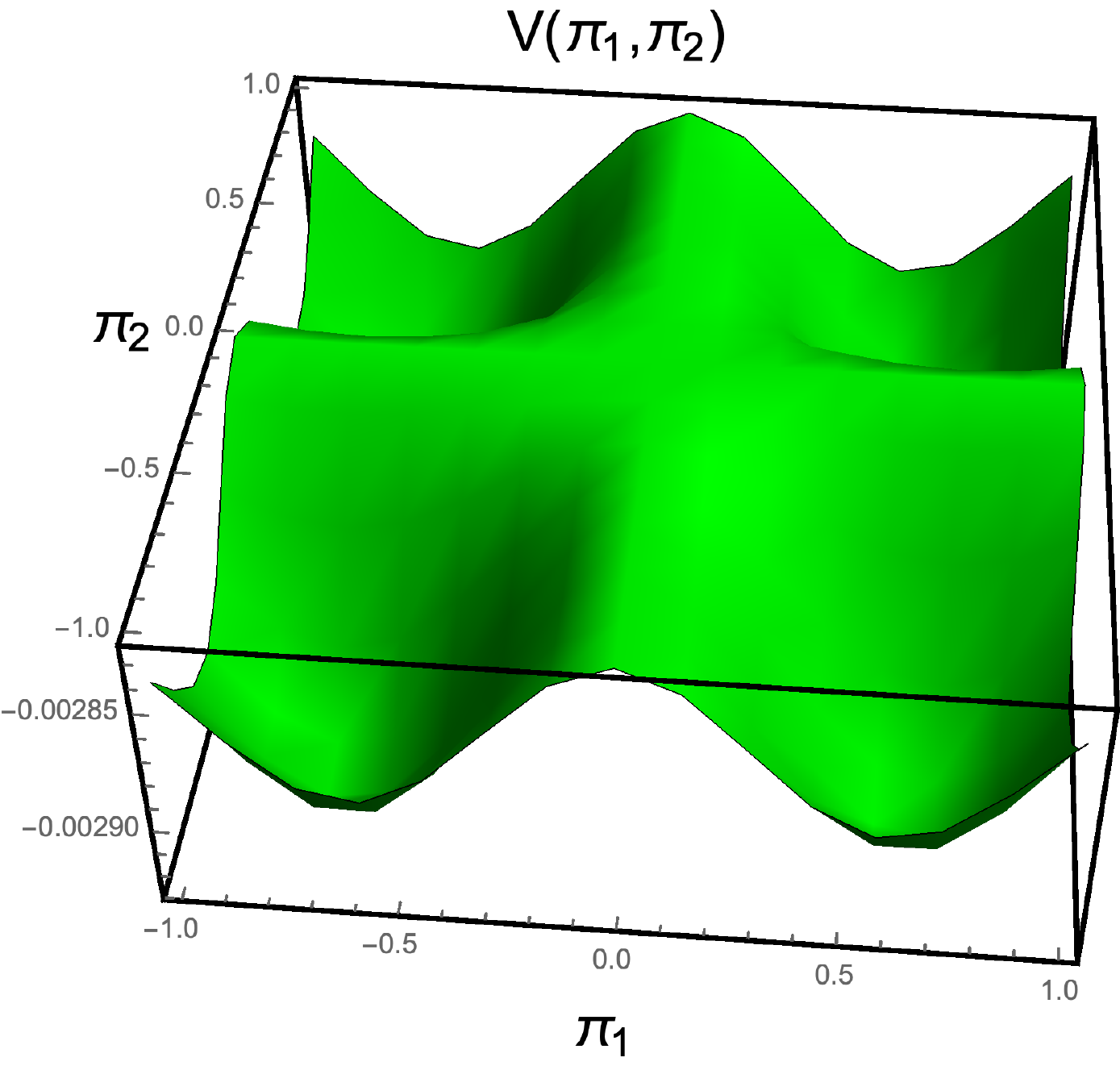}
    \caption{One loop potential for the pions in arbitrary units with $m_\psi=0$, $y=0.5$, $f=1$ and $\lambda=1,\mu=1$.}
    \label{Fig: potential}
\end{figure}

However, the discrete $A_4$ symmetry does not entirely prevent the pions from acquiring a potential.  The one loop Coleman Weinberg potential (in $\overline{\text{MS}}$) \cite{PhysRevD.7.1888,Martin:2001vx} for the pions generated by the fermions is given by
\begin{equation}
\begin{split} \label{Eq: 1-loop}
V_{\text{1 loop, fermions}}&=-\frac{1}{16\pi^2}\text{Tr}\left((M\cdot M^T)^2\left[\ln\left(\frac{M\cdot M^T}{\mu^2}\right)-\frac{3}{2}\right]\right)
\end{split}
\end{equation}
where $\mu$ is the renormalization scale. The logarithmic piece breaks the $SO(3)$ symmetry and generates an effective potential for the pions.  This potential is plotted in Fig.~\ref{Fig: potential}.

The one loop potential is flat
%\footnote{The one-loop potential is flat at the origin for the following reason. For $m_\psi=0$, $\phi\rightarrow -\phi$ is a valid symmetry of the theory which only allows even functions in effective potential. The properties of $A_4$ group strongly restricts the form of effective potetial, $V(\phi)=V(\phi^T\phi,\phi_1^4+\phi_2^4+\phi_3^4)$ which is flat at $\begin{pmatrix}0&0&f\end{pmatrix}$. For $m_\psi\neq 0$, the effective potential, in Fig~\ref{fig:oddonelooppot}, is no longer flat at the origin as odd functions are allowed.} 
along the lines $\pi_1 = 0$ or $\pi_2 = 0$ (this is a one-loop accident as the yukawa coupling is proportional to $\pi_1 \pi_2$) and has 8 degenerate minima that obey $\pi_1= \pm \pi_2$.
The effective potential gives a vev to the pions so that the vev of the scalar $\phi$ is stabilized around any of the eight vacua $ \frac{f}{\sqrt{3}}\begin{pmatrix} \pm1&\pm1&\pm1\end{pmatrix}$ (only four of the eight solutions are shown in Fig.~\ref{Fig: potential}). The mass of the pions in these new vacua are parametrically smaller than the mass of the radial mode.

\begin{figure}[t]
  \centering
  \subfloat{\includegraphics[width=0.45\textwidth]{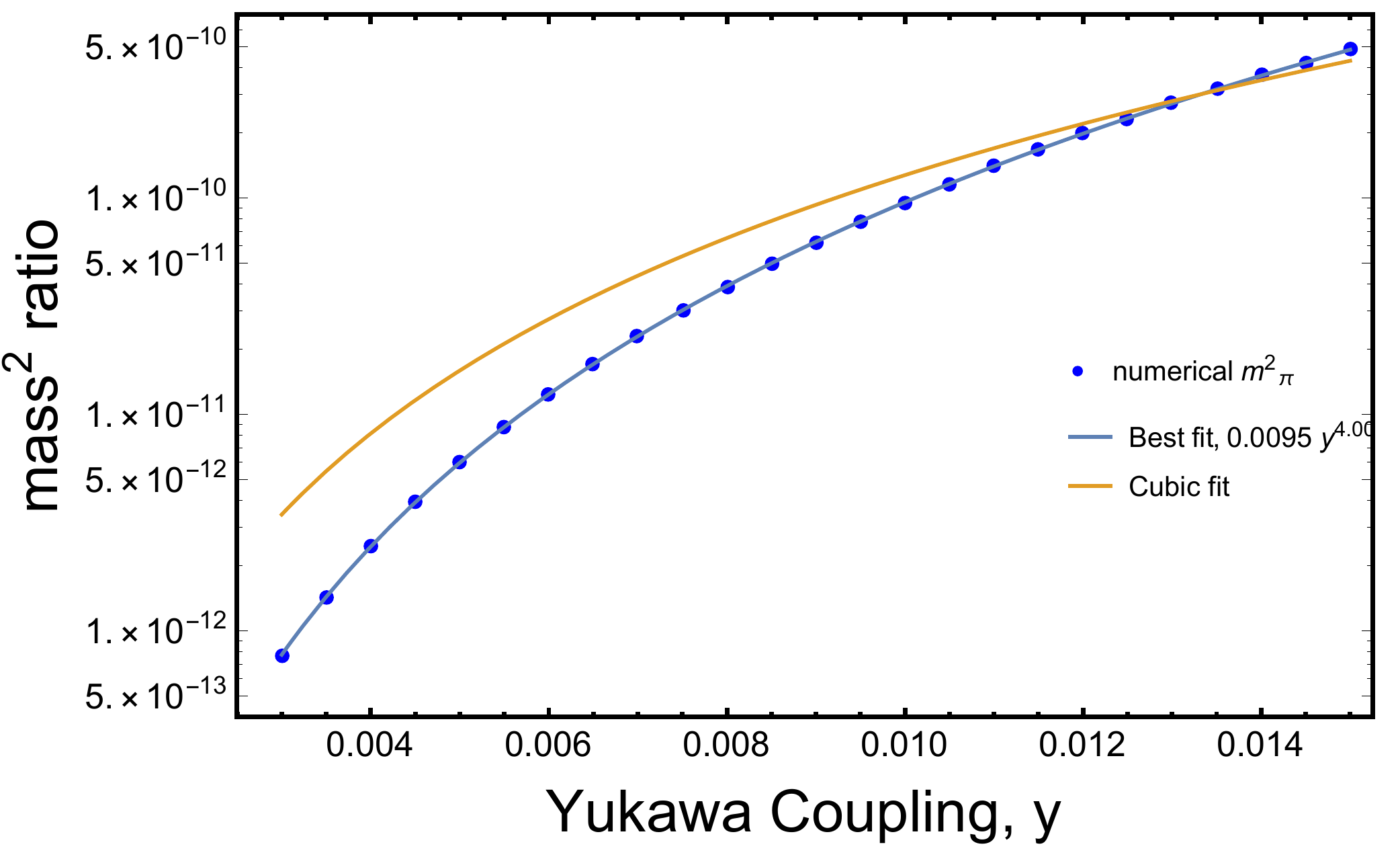}\label{fig:f1}}
  \hfill
  \subfloat{\includegraphics[width=0.45\textwidth]{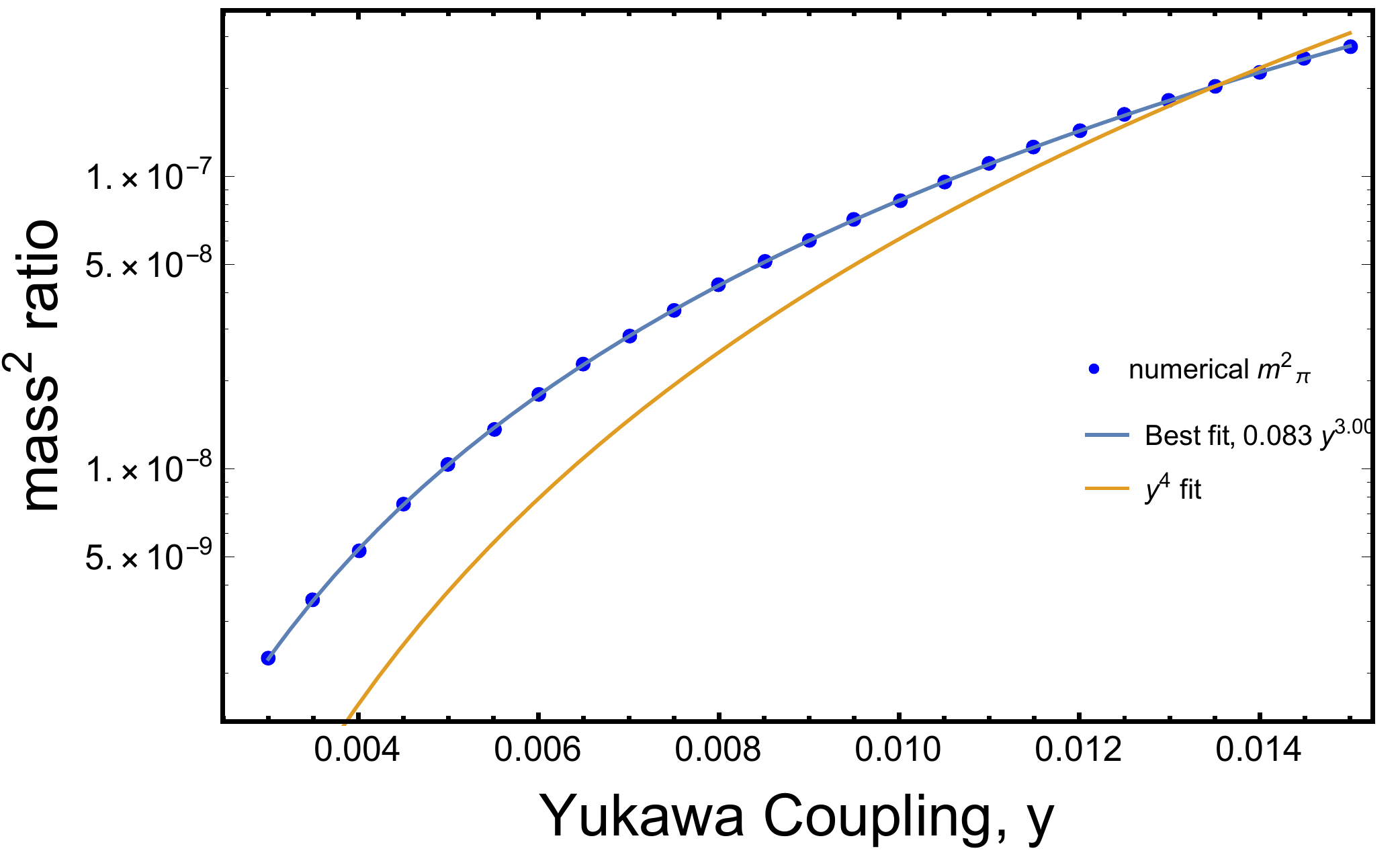}
  \label{fig:f2}}
  \caption{The ratio of the square of the heavier discrete NGBs' mass to the square of the radial mode's mass.  For simplicity we take $f^2=\frac{1}{2}$, $\lambda=1,\mu=1$. In the left panel, we have assumed $m_{\psi}=0$. Under this condition, $\phi\rightarrow -\phi$ is a valid symmetry which prohibits odd functions of $\phi$ in the effective potential.  The mass of the pions comes from a term that scales as $\phi^4$, hence $m_{\pi}^2\propto y^4$. In the right panel, the fermion has a non-zero vector-like mass, $m_{\psi}=1$. The $Z_2$ symmetry is no longer present and pions get a mass from the $\phi^3$ term, and hence $m_{\pi}^2\propto y^3$. }
  \label{Fig: mass suppression}
\end{figure}

In Fig.~\ref{Fig: mass suppression}, we plot the ratio of the numerical values of $m_\pi^2$ to the mass squared of the radial mode.  The pions' mass is suppressed by different powers of Yukawa depending on the presence of or absence of a vector-like fermion mass.  Depending on if there is a vector-like mass, $m_\pi^2$ scales as either $y^3$ or $y^4$.
%Although the argument presented only applies for the particular symmetry, in this case $A_4$, there exists a difference general to all discrete symmetries.\\

In the presence of a large vector-like mass for the fermions, the effective potential becomes analytic and can be expanded in terms of the expansion parameter $\big(\frac{yf}{m_{\psi}}\big)$ .  Schematically, the potential is of the form
\begin{equation}
\begin{split}
V_{A_4}(\phi)&=m_{\psi}^4\sum_{n} C_n \bigg(\frac{y\phi}{m_{\psi}}\bigg)^n 
%\\ \implies V_{A_4}(\pi)&=m_{\psi}^4\sum_{n=3}\bigg(\frac{yf}{m_{\psi}}\bigg)^n \bigg(\sum_{2\leq k\leq n}\frac{\pi_{i_1}}{f}\frac{\pi_{i_2}}{f}\dots\frac{\pi_{i_k}}{f}\bigg)
\end{split}
\end{equation}
As the quadratic part of the scalar potential preserves $SO(3)$, the first non-constant contribution occurs at the third order in the power series expansion.  Since the Yukawa generated potential is of the form $f(y\phi)$, the leading order term in the potential is suppressed by $y^3$ and the the pion mass squared is suppressed by at least the same power of Yukawa coupling.  
In the case under consideration, Eq.~\ref{Eq: 1-loop} can be expanded into the form
\begin{equation}
    \begin{split}
        &V_{\text{1 loop, fermions}}=\mathcal{C}_1+\mathcal{C}_2\phi^T\phi -\frac{ y ^3 m_{\psi} \phi_1 \phi_2 \phi_3 \left(3 \ln \left(\frac{m_{\psi}}{\mu}\right)+1\right)}{\pi^2} +O((\lambda\phi)^4)\\
        &=\mathcal{C}+\frac{ y^3 fm_{\psi} \left(3 \ln \left(\frac{m_{\psi}}{\mu}\right)+1\right)}{ \sqrt{3}\pi ^2  }(\pi_1+\pi_2)^2+\frac{2 y^3 fm_{\psi} \left(3 \ln \left(\frac{m_{\psi}}{\mu}\right)+1\right)}{ 3\sqrt{3}\pi ^2 \cos ^{-1}\left(\frac{1}{\sqrt{3}}\right)^2 }(\pi_1-\pi_2)^2+O(\pi^3)
    \end{split}
\end{equation}
for some constants $\mathcal{C}$, $\mathcal{C}_1$ and $\mathcal{C}_2$. Additionally, in the second line we have expanded the potential around its minimum.  In the next section we will see that in other discrete groups the effective potential is even more suppressed.
On the other hand, in the absence of a fermion mass, $\phi$ has a $\phi \rightarrow -\phi$ symmetry and the leading order term which breaks $A_4$ is of the form $\phi^4$ leading to the scaling $m_\pi^2 \sim y^4$.  However, there is a more dangerous fact hiding in this limit, the effective potential is non-analytic.  The non-analytic part of the potential is of the form
\begin{equation}
V(\phi) \sim \sum_{k} (y \phi)^4  \ln(\frac{y^k\phi^k}{\Lambda^k})  \sim \sum_{k} (y \phi)^4  \ln(\phi^k) .
\end{equation}
The mass generated by this logarithmic part of the potential will only ever be suppressed by $y^4$ regardless of how high in $k$ one must go to obtain a non-zero potential for the pions.  Thus the mass is only guaranteed to be quadratically suppressed in the Yukawa coupling.  This is in complete analogy with collective symmetry breaking models where naive counting will lead one to expect mass terms proportional to $m^2_\pi \sim \prod_i y_i$ but the 1-loop result can be instead proportional to $m_\pi^2 \sim \left ( \prod_i y_i \right)^{(4/N)}$~\cite{ArkaniHamed:2001nc}.

%%%%%%%%%%%%%%%%%%%%%%%%%%%%%  Section 3     %%%%%%%%%%%%%%%%%%%%%%%%%%%%%%%%%%%%%%%%%%%%%%
\section{Analysis using Invariant Theory} \label{Sec: invariance}

Our example of the tetrahedral group illustrates that discrete symmetries partially protect the Goldstones' mass from radiative corrections.  In this section, we show how given a representation of a non-abelian discrete symmetry, one can use the results of invariant theory to calculate how suppressed the mass term should be.

A simple example is useful in obtaining an intuitive picture of the general discussion.  As before, our example will consist of an $A_4$ discrete symmetry with a scalar in the triplet representation.  The starting point involves the invariant polynomials of the triplet representation  
\begin{equation}
\mathcal{I}_2(\phi)=\phi^T\phi\quad\mathcal{I}_3(\phi)=\phi_1\phi_2\phi_3\quad\mathcal{I}_4(\phi)=\sum_i\phi_i^4 .
\end{equation}
The invariant polynomials $\mathcal{I}$ are polynomial functions of $\phi$ that are invariant under $A_4$.  Surprisingly, all $A_4$ invariant functions of $\phi$ can be expressed as a function of just these three invariant polynomials.  For example
\begin{equation}
    \phi^6=\alpha_1 (\mathcal{I}_2(\phi))^3+\alpha_2(\mathcal{I}_3(\phi))^2+\alpha_3 \mathcal{I}_2(\phi)\mathcal{I}_4(\phi)
\end{equation}
for some real numbers $\alpha_i$ that depend on how one contracts the unspecified $A_4$ indices.

Since the potential of $\phi$ respects the underlying symmetry, the effective potential is necessarily of the form
\begin{equation}
V(\phi)=f( \mathcal{I}_2(\phi), \mathcal{I}_3(\phi), \mathcal{I}_4(\phi)).
\end{equation}
Note that $\mathcal{I}_2(\phi)$ is $SO(3)$ invariant and that the first $SO(3)$ non-invariant operator appears at the $\phi^3$ level.  Thus, we can compute how suppressed the potential for the discrete NGBs of the triplet representation of $A_4$ must be by simply finding the first invariant polynomial that does not respect an accidental continuous symmetry.  In this case, the mass term appears at order $\phi^3$.

%For the triplet representation of the tetrahedral group, the basis of the invariant ring has only three polynomials. They are
%\begin{equation}
%\mathcal{I}_2(\phi)=\phi^T\phi\quad\mathcal{I}_3(\phi)=\phi_1\phi_2\phi_3\quad\mathcal{I}_4(\phi)=\sum_i\phi_i^4\
%\end{equation}
%The only possible $6$th degree invariant is a suitable product of these three basic invariants
%\begin{equation}
%    \mathcal{I}_6(\phi)=\alpha_1 (\mathcal{I}_2(\phi))^3+\alpha_2(\mathcal{I}_3(\phi))^2+\alpha_3 \mathcal{I}_2(\phi)\mathcal{I}_4(\phi)
%\end{equation}
%where $\alpha_i$s are real numbers.

We now generalize the previous discussions to all representations of any discrete symmetry group.
Again, the starting point of our analysis are the invariant polynomials of a symmetry group.  As before, given a M dimensional representation $\phi_1, \cdots , \phi_M$, the invariant polynomials are sums of products of the $\phi$s which are invariant under the discrete symmetry.  We will use the convention $\mathcal{I}_n$ to denote an invariant polynomial of degree n.
The set of invariant polynomials is called the invariant ring. The elements of the invariant ring of any discrete group can be expressed as a polynomial of finite number of algebraically independent functions, therefore, the invariant ring of a discrete group is `finitely generated' \cite{Merle_2012}.  This situation is analogous to the fundamental representation of the Orthogonal group, for which all invariant functions can be expressed as polynomials of $\phi^T \phi$.

There is thus a simple recipe for determining when the potential for a discrete NGB is non-zero.  
Given an M dimensional representation, its maximal symmetry group is $SO(M)$ or $SU(M)$.  Invariant polynomial of lowest degree typically respect these accidental symmetries.
Simply look up the invariant polynomials of the representation and group of interest, and find the polynomial of lowest degree that breaks the accidental global symmetry.  The degree of this polynomial gives the degree of suppression of the discrete NGB potential.

The ``ideal" situation is if the invariant polynomial that breaks the accidental global symmetry is of very high degree.  The easiest way to enforce this condition is if there are not many low dimensional invariant polynomials.  There are several mathematical proofs that are useful when searching for such a representation.  
Combined, these theorems will tell us that when looking for highly suppressed potentials, it is best to look for small representations of groups that have a large number of elements.  The most extreme example is $\ZN$ which has N elements but only has a one-dimensional representation.

The first mathematical proof is a remarkable theorem \cite{Hilbert1890,Noether1915,article, zbMATH02672848} that guarantees that the number of algebraically independent invariant functions equals the dimension of the representation.  The second theorem is the following \cite{coxeter1951,shephard_todd_1954}: If $\mathcal{H}$ is a finite subgroup generated by reflections~\footnote{A reflection is a diagonalizable non-identity linear isomorphism of finite order that keeps all the points on a hyperplane fixed. The matrix representation of a reflection has all of its eigenvalues equal to 1 except for a single eigenvalue whose value is the $m$th root of unity where $m$ is the order of the reflection.} of a unitary group of $n$ variables, then $\mathcal{H}$ posses $n$ algebraically independent invariant forms $\mathcal{I}_{m_1}$, $\mathcal{I}_{m_2}$, $\ldots$ , $\mathcal{I}_{m_n}$ with degrees $m_1$, $m_2$, $\ldots$ , $m_n$ such that
\begin{equation}
\prod^n_{i=1}m_i=g
\end{equation}
where $g$ is the number of elements in the group. We can readily check the way the theorems apply to $Z_N$, since $Z_N$ is a reflection generated subgroup of $U(1)$.  $Z_N$ has a dimension one representation that then only has one invariant polynomial. As the number of elements in $Z_N$ is $N$, the degree of the polynomial is forced to be $N$, hence the only allowed function is $\phi^N$. As an example of a non-Abelian group, we can check $A_5\times Z_2$. $A_5$ is isomorphic to the proper rotations of an Icosahedron. Being a subgroup of rotation, $A_5$ doesn't include reflections, but $A_5\times Z_2$ does. It has a total 120 elements. Its three dimensional representation has three basic invariants. Other than the familiar degree 2 invariant $\phi^T\phi$, two other invariants have degrees 6 and 10, making $\prod^3_{i=1}m_i=2\cdot 6\cdot 10 =120$.

The combination of the theorems forces the following conclusion. If one holds the dimension of the representation fixed, i.e. the number of $m_i$'s, but have very large $m_i$'s, the only way to achieve this is to increase $g$.
% To formalize this statement, we can borrow from classic results of invariant theory of finite subgroups of Orthogonal groups \cite{coxeter1951} (unitary groups \cite{shephard_todd_1954}) generated by reflections.
Finally we conclude by listing the degrees of the basic polynomial invariants for a few familiar groups.  $T'$ illustrates an important subtlety.  The two dimensional representation is complex, however the familiar second degree invariant $\phi^* \phi$ is not a polynomial in $\phi$ in strict mathematical sense and thus is not constrained by these considerations.

\begin{table}[H]
    \centering
    \resizebox{0.55\columnwidth}{!}{%
    \begin{tabular}{ |m{1.3cm} | m{1.0cm} | m{2cm} | m{2.0cm} | m{2cm} | }
    \hline
    Group & $n$ & Parent Lie Group & Number of Elements & Degree of Invariants \\ 
    \hline
    $A_4$ & 3 & $SO(3)$ & 12 & 2,3,4 \\ 
    \hline
    $S_4$ & 3 & $O(3)$ & 24 & 2,3,4 \\ 
    \hline
    $T'$ & 2 & $SU(2)$ & 24 & 6,8 \\ 
    \hline
     $S_4$ & 3 & SO(3) & 24 & 2,4,6\\
    \hline
    $S_4 \times Z_2$ & 3 & O(3) & 48 & 2,4,6\\
    \hline
    $A_5 \times Z_2$ & 3 & O(3) & 120 & 2,6,10\\
    \hline
    \end{tabular}%
    }
    \caption{Degree of Polynomial invariants for a few groups.  $n$ is the dimension of the representation under consideration.  For subgroups of SO(3) or SU(2), the theorem doesn't apply.  First three groups are isomorphic to tetrahedron, next two to Cube or Ocathedron. $A_5\times Z_2 $ is isomorphic to Icosahedron.}
    \label{tab:my_label}
\end{table}

%%%%%%%%%%%%%%%%%%%%%%%%%%%%%  Section 4     %%%%%%%%%%%%%%%%%%%%%%%%%%%%%%%%%%%%%%%%%%%%%%
\section{Exchange representation} \label{Sec: exchange}

\begin{figure}[H] 
    \centering
    \includegraphics[width=0.45\textwidth]{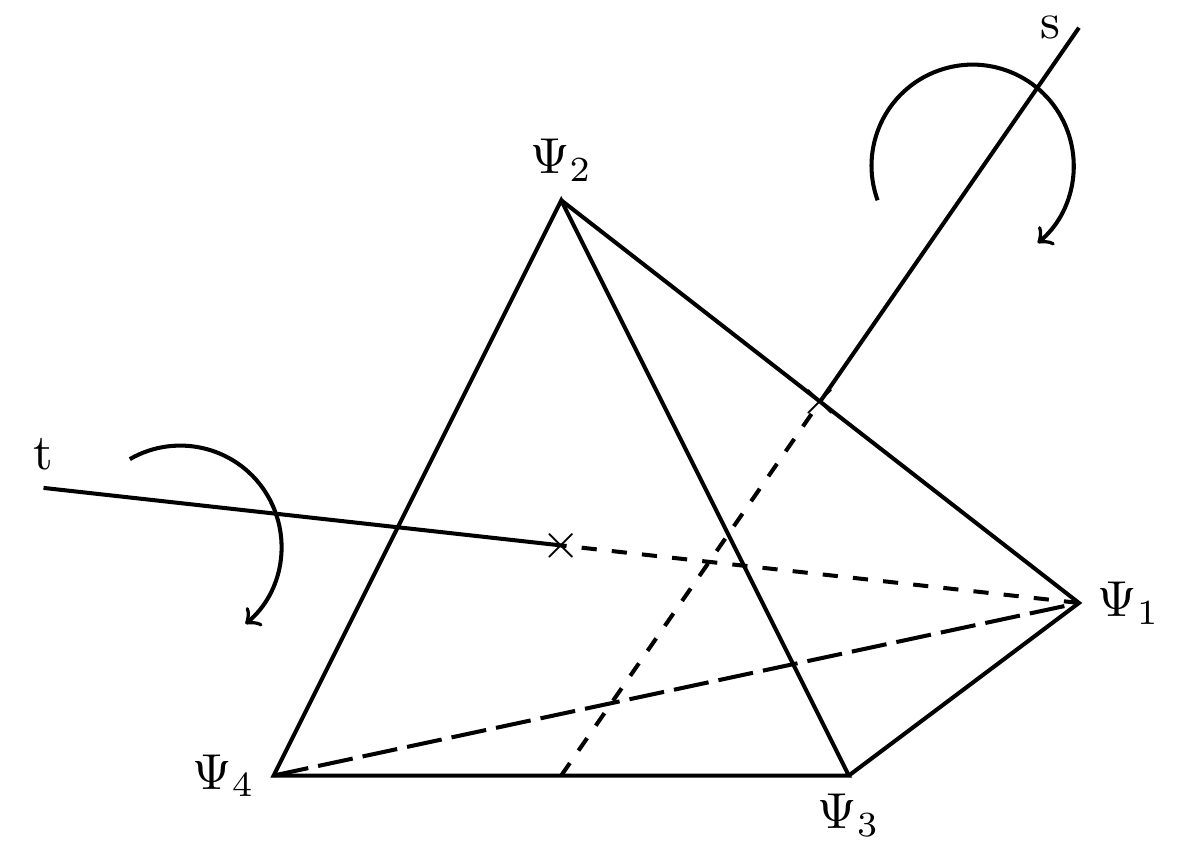}
    \caption{The group of proper rotations of a tetrahedron is isomorphic to the group of even permutations of four elements, $A_4$. $A_4$ can be succinctly parameterized by two rotations, $s$ and $t$, which satisfy $s^2=t^3=(st)^3=e$. The span of these two rotations covers $A_4$. The four fermions occupying the vertices exchange under $A_4$ as shown in Eq.~\ref{eq: st}.
    %$s: \Psi_1\leftrightarrow \Psi_2, \Psi_3\leftrightarrow \Psi_4$, $t:\Psi_2\mapsto\Psi_3\mapsto\Psi_4\mapsto\Psi_2, \Psi_1\mapsto\Psi_1$.
    }
    \label{Fig: tetrahedron}
\end{figure}

Linear representations require all of the particles in the representation to have the same gauge quantum numbers.
The reason for this is that gauge groups are usually uncharged under the transforming symmetries.  
A simple example of this in the Standard Model is that all three colors of the left handed up quark have the exact same gauge quantum numbers.
Aside from linear representations, discrete symmetries are useful because they have exchange representations.
%Gauge symmetries play an important role in describing nature. For example, the standard model is well described by $SU(3)\otimes SU(2)\otimes U(1)$ gauge symmetry. But discrete symmetries, by being discrete, can't be gauged in the customary fashion. One simple solution would be to consider direct product of gauge groups and discrete groups. The gauged charged particles will then transform trivially under the discrete group.\\
This new representation allows for the new possibility that the gauge groups are in the exchange representation of the discrete symmetry, where copies of the gauge group transform into each other under the action of the discrete symmetry~\footnote{This situation can be useful in theories such as Twin Higgs, where making partners charged under different gauge symmetries drastically changes the phenomenology.}. For the example of the Tetrahedral group, we can imagine four fields with their own different gauge groups representing the vertices. The fields along with their gauge sectors interchange among themselves under the group action.  A pictoral representation of this scenario is shown in Fig.~\ref{Fig: tetrahedron}.  

To elaborate on the example of the Tetrahedron, let us consider four fermions $\Psi_j$ with $1\leq j \leq 4$ representing the four vertices of a tetrahedron in \figureautorefname{ \ref{Fig: tetrahedron}} charged under $SU(N)_{{j,1\leq j \leq 4}}$ gauge symmetries . \\
\begin{equation}
    \begin{split}
        \mathcal{L}_{\text{kin}}&=\sum_j  \mathcal{L}_{\text{kin},j}\\
        \mathcal{L}_{\text{kin},j}&=\overline{\Psi}_j i\gamma^{\mu}\big(\partial_{\mu}-igA_{\mu,j}^a\tau_a\big)\Psi_j
    \end{split}
\end{equation}
Under the action of $A_4$, the fermions and the gauge fields get exchanged.
\begin{equation} \label{eq: st}
    \begin{split}
        s: &(\Psi_1,A_{\mu,1}^a)\leftrightarrow (\Psi_2,A_{\mu,2}^a),\ \ \ (\Psi_3,A_{\mu,3}^a)\leftrightarrow (\Psi_4,A_{\mu,4}^a)\\
        t:&(\Psi_2,A_{\mu,2}^a)\mapsto(\Psi_3,A_{\mu,3}^a)\mapsto(\Psi_4,A_{\mu,4}^a)\mapsto(\Psi_2,A_{\mu,2}^a),\ \ \ (\Psi_1,A_{\mu,1}^a)\mapsto(\Psi_1,A_{\mu,1}^a)
    \end{split}
\end{equation}
Since the sectors are related by exchange symmetry, the fermions are all charged under different gauge groups despite being in the same representation of $A_4$.

There are several ways to couple a scalar to fermions in an exchange representation.  If the scalar is also in the exchange representation, then it is trivial to add a new scalar per site.  Instead, we will focus on the case where the scalar is in a linear representation and thus has the feature mentioned in the previous sections of having a suppressed potential.
The key to coupling a linear and an exchange representation is to realize that the exchange representation can be decomposed into linear representations.
The $A_4$ reducible exchange representation of four fermions can be decomposed into a singlet and a triplet \cite{Ishimori:2010au}.
\begin{equation}
    \begin{pmatrix}
    \overline{\Psi_1}\Psi_1\\ \overline{\Psi_2}\Psi_2\\\overline{\Psi_3}\Psi_3\\\overline{\Psi_4}\Psi_4 \end{pmatrix}_\mathbf{4}^{\text{exchange}}=\begin{pmatrix} \overline{\Psi_1}\Psi_1+\overline{\Psi_2}\Psi_2+ \overline{\Psi_3}\Psi_3+ \overline{\Psi_4}\Psi_4\end{pmatrix}_\mathbf{1}\oplus \begin{pmatrix}\overline{\Psi_1}\Psi_1+\overline{\Psi_2}\Psi_2- \overline{\Psi_3}\Psi_3- \overline{\Psi_4}\Psi_4 \\ \overline{\Psi_1}\Psi_1-\overline{\Psi_2}\Psi_2+ \overline{\Psi_3}\Psi_3- \overline{\Psi_4}\Psi_4 \\ \overline{\Psi_1}\Psi_1-\overline{\Psi_2}\Psi_2- \overline{\Psi_3}\Psi_3+ \overline{\Psi_4}\Psi_4 \end{pmatrix}_\mathbf{3}
\end{equation}
A triplet scalar couples to the irriducible triplet of the exchange representation by the yukawa interaction
\begin{equation}
    \mathcal{L}_{\text{int}}=y\begin{pmatrix}\overline{\Psi_1}\Psi_1+\overline{\Psi_2}\Psi_2- \overline{\Psi_3}\Psi_3- \overline{\Psi_4}\Psi_4 \\ \overline{\Psi_1}\Psi_1-\overline{\Psi_2}\Psi_2+ \overline{\Psi_3}\Psi_3- \overline{\Psi_4}\Psi_4 \\ \overline{\Psi_1}\Psi_1-\overline{\Psi_2}\Psi_2- \overline{\Psi_3}\Psi_3+ \overline{\Psi_4}\Psi_4 \end{pmatrix}\cdot\begin{pmatrix}\phi_1\\ \phi_2 \\ \phi_3 \end{pmatrix}
\end{equation}
The form of the interaction carries the signature of the underlying symmetry, in this case the $A_4$ asymmetry.  Thus, despite the rather strange appearance of the interaction, the pion masses will again be highly suppressed, in this case by either $y^3$($y^4$) if there is (is not) a vector-like mass.

We can readily apply all the tools mentioned above once we identify all discrete groups which allow for an exchange representation. The amazing feature of discrete groups is that all of them do. The easiest way to construct one is to observe that the group elements exchange among themselves under the action of the group. So every discrete group has an exchange representation with dimensions equal to the number of elements in the group itself.  This is a consequence of Cayley's Theorem \cite{doi:10.1080/14786445408647421,ref000419368}, which states that every group $G$ is isomorphic to a subgroup of symmetric group acting on $G$.  

The case can be best illustrated with a cyclic group like $Z_N$. $Z_N$ has a single generator $a$ and the N elements of the group are simply $(e,a^1,a^2,\dots, a^{N-1})$.  These N elements can be converted into an N dimensional exchange representation using $N$ scalar fields $(\phi_1,\phi_2,\dots, \phi_N)$ that are permuted under the action of the group as
\begin{equation}
    a^k \begin{pmatrix}\phi_1\\ \phi_2\\ \vdots\\\phi_{N-k}\\\vdots\\ \phi_N\end{pmatrix}=\begin{pmatrix}\phi_{k+1}\\ \phi_{k+2}\\\vdots\\\phi_N\\\vdots\\\phi_{k}\end{pmatrix}
\end{equation}
However, the $N$ dimensional exchange representation is reducible since $Z_N$, being Abelian, only allows for one dimensional irreducible representations (irreps). The singlet can be identified as the linear combination $(\phi_1+\phi_2+\cdots +\phi_N)$ and a general irreducible representation is furnished by
\begin{equation}
    e^{\frac{i2k\pi}{N}}:(\phi_1+e^{\frac{i2k\pi}{N}}\phi_2+\cdots+e^{\frac{i(N-1)2k\pi}{N}}\phi_N),\ \ \ \ 0\leq k\leq (N-1)
\end{equation}
The procedure can be generalized to non-Abelian discrete groups which allows for higher dimensional irreps.
%\SD{I had to remove the whole part corresponding to the subgroups. I realized that all the schemes I could think of about restriction of exchange representation on a subgroup was actually removing all vestige of the larger parent group. It's then just exchange representation of a smaller group, not of any subgroup of a larger group.}\\
%For smaller exchange representations, we can again use Cayley's theorem restricted on subgroups. Since any subgroup furnishes another exchange representation, existence of smaller exchange representation depends upon few details of the group. In particular, "Converse Lagrange Theorem (CLT) group"s have exchange representation with dimensions equal to all the divisors of the order of the group. For example, the symmetric group of four elements, $S_4$, isomorphic to the proper rotations of a cube is a CLT group. 

The exchange representations generated from Cayley's theorem can be too large to be convenient. For example, the now familiar $A_4$ group has a $12$ dimensional exchange representation. However, in many cases, there exist smaller exchange representations which also have a simple geometrical interpretation. For example, $A_4$ has four and six dimensional exchange representations which exchange vertices~\footnote{The representations exchanging four vertices and four faces are equivalent.} and edges respectively of a regular tetrahedron. For $A_5$, the group is isomorphic to proper rotations of an Icosahedron, representations with dimensions of twelve, twenty and thirty that exchange its vertices, faces and edges respectively.\\

%%%%%%%%%%%%%%%%%%%%%%%%%%%%%%%%%%%%%%%%%%%%%%%%%%%%%%%%%%%%%%%%%%%%%%%%%%%%%%%%%%%%%%%%%%%%%%%%%%%%%%%%%%%%%%%%%

%%%%%%%%%%%%%%%%%%%%%%%%%%  Section 5    %%%%%%%%%%%%%%%%%%%%%%%%%%%%%%%%%%%%%%%%%%%%%%
\section{Conclusions} \label{Sec: conclusion}

In this paper we initiated a study of the pseudo-Nambu Goldstone bosons of discrete symmetries. NGBs of discrete symmetries feature many interesting properties.  On one hand, discrete NGBs can have large shift symmetry breaking Yukawa interactions.  On the other hand, discrete NGBs have suppressed potentials, e.g. the quadratically divergent contribution to their potential is not present.

In order to determine how suppressed the potential of a discrete NGB is, first rewrite the non-linear realization in terms of a linear realization.  Then find the lowest dimensional operator that breaks the accidental continuous symmetry and gives mass to the discrete NGB.  In many cases, there is not a renormalizable operator that can be written down that gives mass to the NGB.
We illustrated this procedure in a toy model example where we gave an $A_4$ theory with a triplet where one could explicitly demonstrate that the potential for the discrete pNGB was suppressed.

Many of these considerations have already been expressed in some form or another mathematically.  One example is the idea of invariant polynomials.  All products of fields can be rewritten in terms of the invariant polynomials (in much the same way that in an $SO$ theory all gauge invariant operators involving the vector $\phi$ are just functions of $\phi^T  \phi$).  By examining the invariant polynomials, one can find at what order the accidental continuous symmetries are broken.

Non-abelian discrete symmetries are exciting and it is somewhat surprising that their Nambu Goldstone bosons have not been considered in detail before.  We have only scratched the surface of their properties.  Perhaps one of the most exciting future directions would be if it were possible to gauge various sub-groups of the accidental symmetries so that one could actually charge discrete NGBs under a gauge symmetry.
Many of the examples of pNGBs that are found experimentally and considered theoretically have gauge quantum numbers.  It would be exciting if the NGBs of discrete symmetries could also have this property.  The suppression of the discrete NGB potential is strongest in vector-like theories that generate analytic potentials.  It would be exciting if the cancellations that occur in chiral theories could be made equally strong.
pNGBs have guided our thinking for a long time and it would be interesting if discrete NGBs change how we approach model building.

%%%%%%%%%%%%%%%%%%%%%%%%%%  Acknowledgement   %%%%%%%%%%%%%%%%%%%%%%%%%%%%%%%%%%%%%%%%%%%%%%
\section*{Acknowledgments} 

The authors thank Prateek Agrawal and Gustavo Marques-Tavares for useful comments on the draft.  This research was supported in part by the NSF under Grant No. PHY-1914480 and by the Maryland Center for Fundamental Physics (MCFP).

%%%%%%%%%%%%%%%%%%%%%%%%%%  Appendix  %%%%%%%%%%%%%%%%%%%%%%%%%%%%%%%%%%%%%%%%%%%%%%
\begin{appendices}

\section{ More details on the $A_4$ invariant potential}

In Sec.~\ref{Sec: explicit example}, we presented the $A_4$ invariant one loop potential with massless fermions in Fig.~\ref{Fig: potential}, which was flat at the origin.  Here we emphasise that this behavior occurs only at the 1-loop level and is specific to the case of massless fermions only.  At 2-loops, the flat directions disappear.

For massive fermions, the broken $Z_2$ symmetry allows odd functions in the effective potential. The total one loop potential now has minima located on the straight lines $\pi_1=\pi_2$ whereas the maxima are on $\pi_1=-\pi_2$.
\begin{figure}[H]
    \centering
    \includegraphics[width=0.45\textwidth]{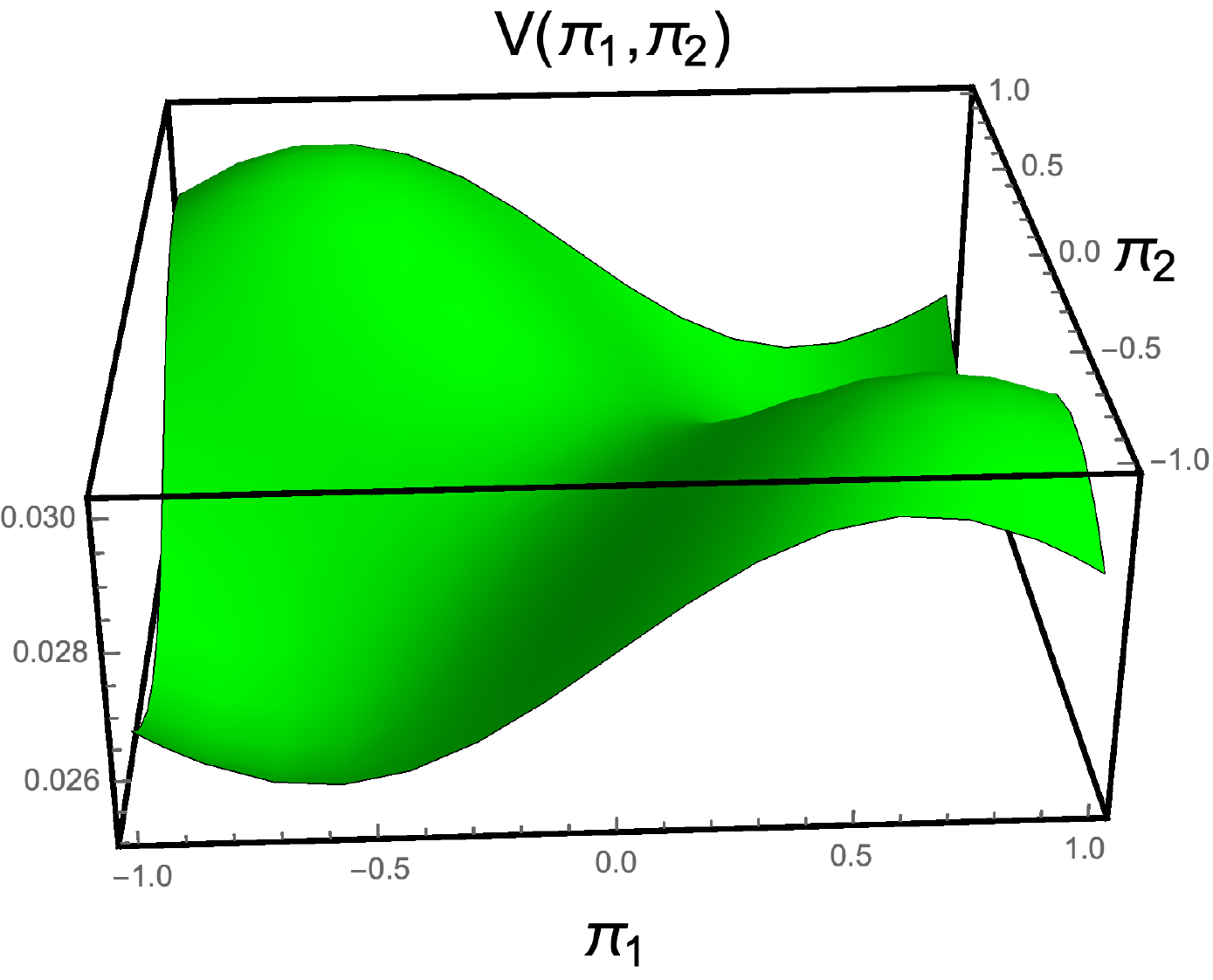}
    \caption{The $A_4$ symmetric one loop potential for $m_{\psi}=1$, $y=0.5$, $f^2=1$, $\lambda=1$, $\mu=1$. The potential is no longer flat at the origin. The mode $\frac{\pi_1-\pi_2}{\sqrt{2}}$ has a regular mass where as $\frac{\pi_1+\pi_2}{\sqrt{2}}$ has a tachyonic mass at $(0,0)$.}
    \label{fig:oddonelooppot}
\end{figure}

%Even in the case of a massless fermion, the flatness of the potential around the origin is the two loop effective potential \cite{Martin:2001vx} has a different qualitative feature as it produces a minima at the origin.
%\begin{figure}[H]
%    \centering
%    \includegraphics[width=0.45\textwidth]{images/two_loop_potential.pdf}
%    \caption{Two loop potential for the pions in arbitrary units. $m_{\psi}=0$, $y=0.1$, $f^2= 1$, $\lambda=1$ and $\mu=1$.}
%\end{figure}

%\begin{figure}[H]
%    \centering
%    \includegraphics[width=0.45\textwidth]{images/two_loop_mass.pdf}
%    \caption{The ratio of Pseudo Goldstones' mass to radial mass. $m_{\psi}=0$, $f^2=\frac{1}{2}$, $\lambda=1$ and $\mu=1$.}
%\end{figure}

\section{Platonic Solids}

The regular convex polyhedrons, which are also known as Platonic solids, are useful examples of discrete symmetries. In three dimensions, there are only five of such objects. In table \ref{tab:Polyhedron}, we present the invariant analysis of the Platonic solids.
\begin{table}[H]
    \centering
    \resizebox{0.55\textwidth}{!}{%
    \begin{tabular}%{|c|c|c|c|c|}
    { | m{2.3cm} | m{0.4cm} | m{1.3cm} | m{0.6cm} | m{4.0cm} | } 
    \hline
    Solid & $\mathcal{R}$ & $\mathcal{S}$  & $|\mathcal{S}|$ & Degree of Invariants \\ 
    \hline
    Tetrahedron & $A_4$ & $S_4$ & 24 & 2,3,4 \\ 
    \hline
    Cube & $S_4$ & $S_4\times Z_2$ & 48 & 2,4,6 \\ 
    \hline
    Octahedron & $S_4$ & $S_4\times Z_2$ & 48 & 2,4,6 \\ 
    \hline
    Icosahedron & $A_5$ &$A_5\times Z_2$ & 120 & 2,6,10 \\ 
    \hline
    Dodecahedron & $A_5$ &$A_5\times Z_2$ & 120 & 2,6,10 \\ 
    \hline
    \end{tabular}%
    }
    \caption{The symmetries of five Platonic solids. $\mathcal{R}$ and $\mathcal{S}$ denote Rotation Group and Symmetry Group respectively.  More colloquially, $\mathcal{R}$ is the discrete group representing the symmetries of the Platonic solid while $\mathcal{S}$ is the double cover.  $|\mathcal{S}|$ denotes the number of elements in the symmetry group.}
    \label{tab:Polyhedron}
\end{table}

The rotation groups of Platonic solids are subgroups of rotation group in three dimensions, $SO(3)$. But they have irreducible representation with dimensions other than 3. Here we present the invariant analysis summary of those representations.
\begin{table}[H]
    \centering
    \resizebox{0.55\textwidth}{!}{%
    \begin{tabular}{ |m{1.3cm} | m{0.5cm} | m{2.0cm} | m{1.5cm} | m{2.0cm} | }
    \hline
    Group & $n$ & Parent Lie Group & Order & Degree of Invariants \\ 
    \hline
    $S_4$ & 2 &$ SU(2)$ & 24 & 2,3 \\ 
    \hline
    $A_5$ & 4 &$ SO(4)$ & 60 & 2,3,4,5 \\
    \hline
    $A_5$ & 5 & $SO(5)$ & 60 & 2,3,3,4,5\\
    \hline
    \end{tabular}%
    }
    \caption{Degree of Polynomial invariants for a few other irreps of the familiar groups.}
    \label{tab:my_label}
\end{table}

Finally, all of the Platonic Solids furnish exchange representations of various discrete groups.  We conclude by demonstrating how these different exchange representations can be decomposed into the standard linear representations.

\subsection{Tetrahedron}
\begin{figure}[H]
    \centering
    \includegraphics[width=0.45\textwidth]{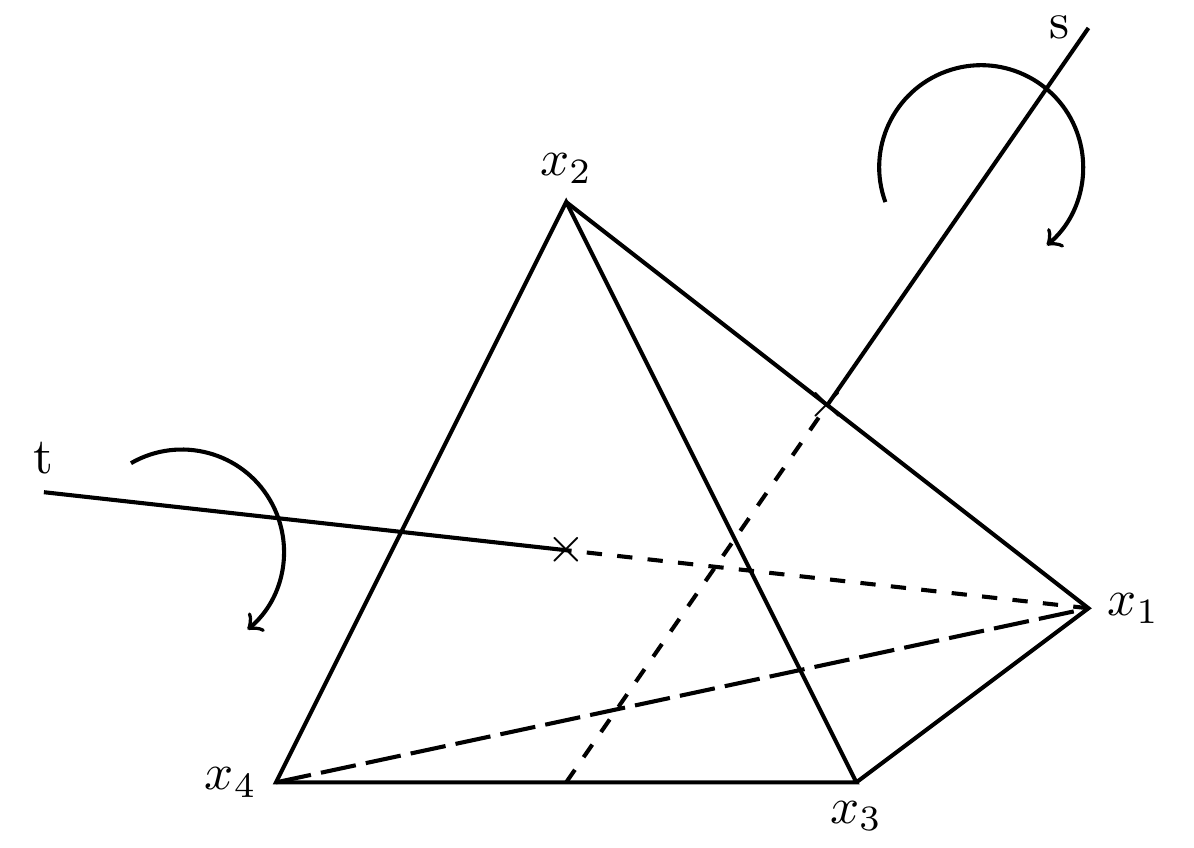}
    \caption{Proper rotations of a regular tetrahedon are isomorphic to $A_4$. $A_4$ can be parameterized be $s^2=t^3=(st)^3=e$. The four vertices exchange under $A_4$ as $s: x_1\leftrightarrow x_2, x_3\leftrightarrow x_4$, $t: x_2\mapsto x_3\mapsto x_4\mapsto x_2, x_1\mapsto x_1$.}
    \label{fig: T}
\end{figure}

The 4 dimensional exchange representation can be decomposed as 
\begin{equation}
    \begin{pmatrix}
    x_1\\ x_2 \\ x_3\\ x_4 \end{pmatrix}_\mathbf{4}^\text{exchange}=\begin{pmatrix} x_1+x_2+ x_3+ x_4\end{pmatrix}_\mathbf{1}\oplus\begin{pmatrix}x_1+x_2- x_3- x_4 \\ x_1-x_2+ x_3- x_4 \\ x_1-x_2- x_3+ x_4 \end{pmatrix}_\mathbf{3} .
\end{equation}

\subsection{Cube}
\begin{figure}[H]
    \centering
    \includegraphics[width=0.45\textwidth]{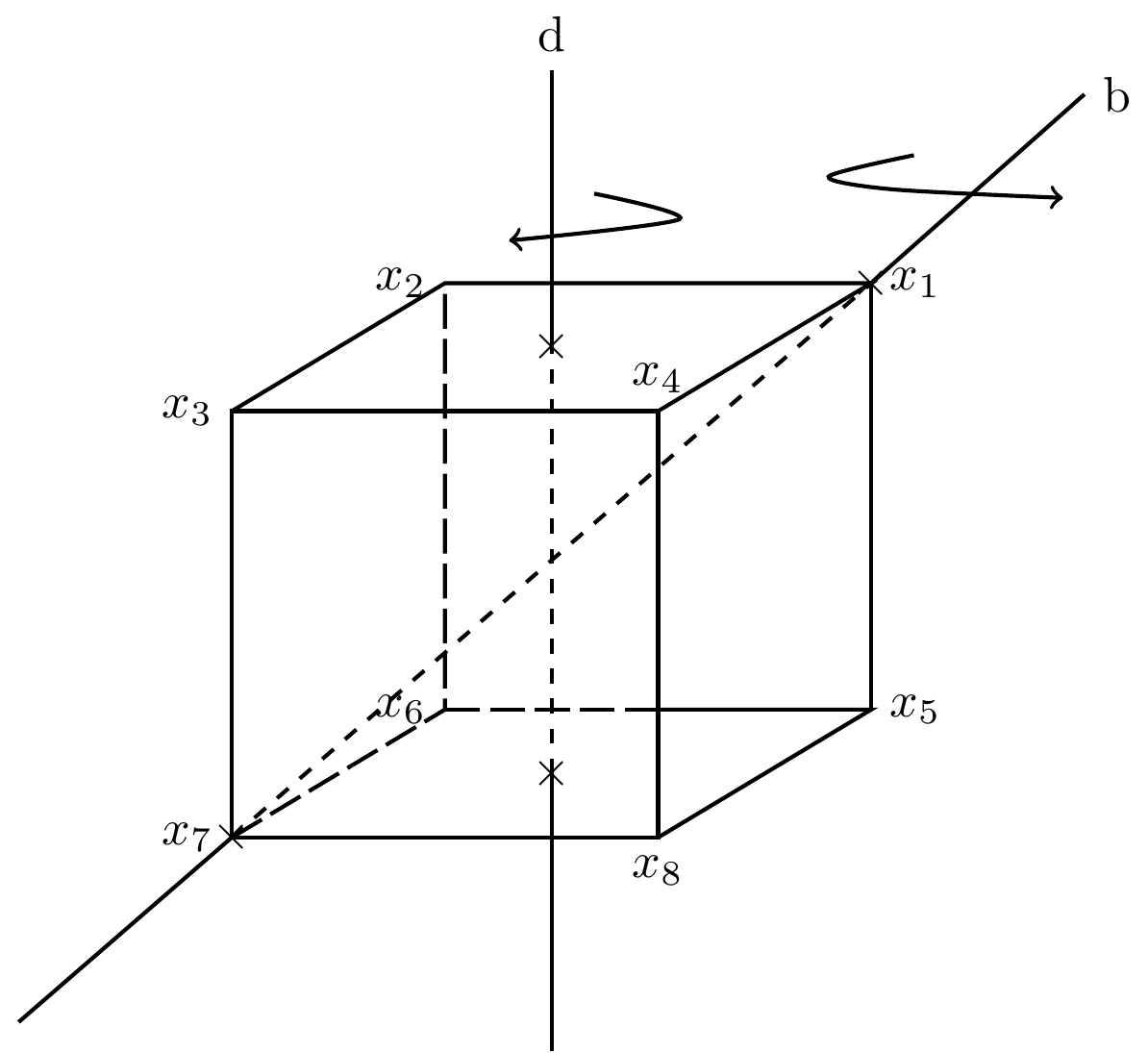}
    \caption{The group of proper rotations of a cube are isomorphic to $S_4$, the group of all possible permutations of four objects. $S_4$ can be parameterized by ~\cite{Ishimori:2010au} $b^3=d^4=e$, $db^2d=b$ and $dbd=bd^2b$. The eight vertices exchange under $S_4$ as $b: x_1 \mapsto x_1, x_7\mapsto x_7, x_2 \mapsto x_4 \mapsto x_5 \mapsto x_2, x_3\mapsto x_8\mapsto x_6 \mapsto x_3$, $d: x_1\mapsto x_4\mapsto x_3\mapsto x_2 \mapsto x_1, x_5\mapsto x_8 \mapsto x_7 \mapsto x_6 \mapsto x_5 $.}
    \label{fig: C}
\end{figure}

The 8 dimensional exchange representation can be decomposed as 
\begin{equation}
    \begin{split}
        \mathbf{8}&=\mathbf{1}\oplus\mathbf{1'}\oplus\mathbf{3}\oplus\mathbf{3'}\\
         \begin{pmatrix}
    x_1\\ x_2 \\ x_3\\ x_4 \\x_5\\x_6\\x_7\\x_8 \end{pmatrix}_\mathbf{8}^\text{exchange}=&\begin{pmatrix} x_1+x_2+ x_3+ x_4+x_5+x_6+x_7+x_8\end{pmatrix}_\mathbf{1}\\
    &\oplus \begin{pmatrix} x_1-x_2+ x_3- x_4-x_5+x_6-x_7+x_8\end{pmatrix}_\mathbf{1'}\\
    & \oplus\begin{pmatrix}x_1- x_2+ x_3- x_4+x_5- x_6+ x_7-x_8 \\ x_1+ x_2- x_3- x_4- x_5- x_6+ x_7+ x_8 \\ x_1-x_2- x_3+ x_4- x_5+ x_6+ x_7- x_8\end{pmatrix}_\mathbf{3}\\
    &\oplus\begin{pmatrix}x_1+ x_2+ x_3+ x_4-x_5- x_6- x_7-x_8 \\ x_1- x_2- x_3+ x_4+ x_5- x_6- x_7+ x_8 \\ x_1+x_2- x_3- x_4+ x_5+ x_6- x_7- x_8\end{pmatrix}_\mathbf{3'} .
    \end{split}
\end{equation}

\subsection{Octahedron}
\begin{figure}[H]
    \centering
    \includegraphics[width=0.45\textwidth]{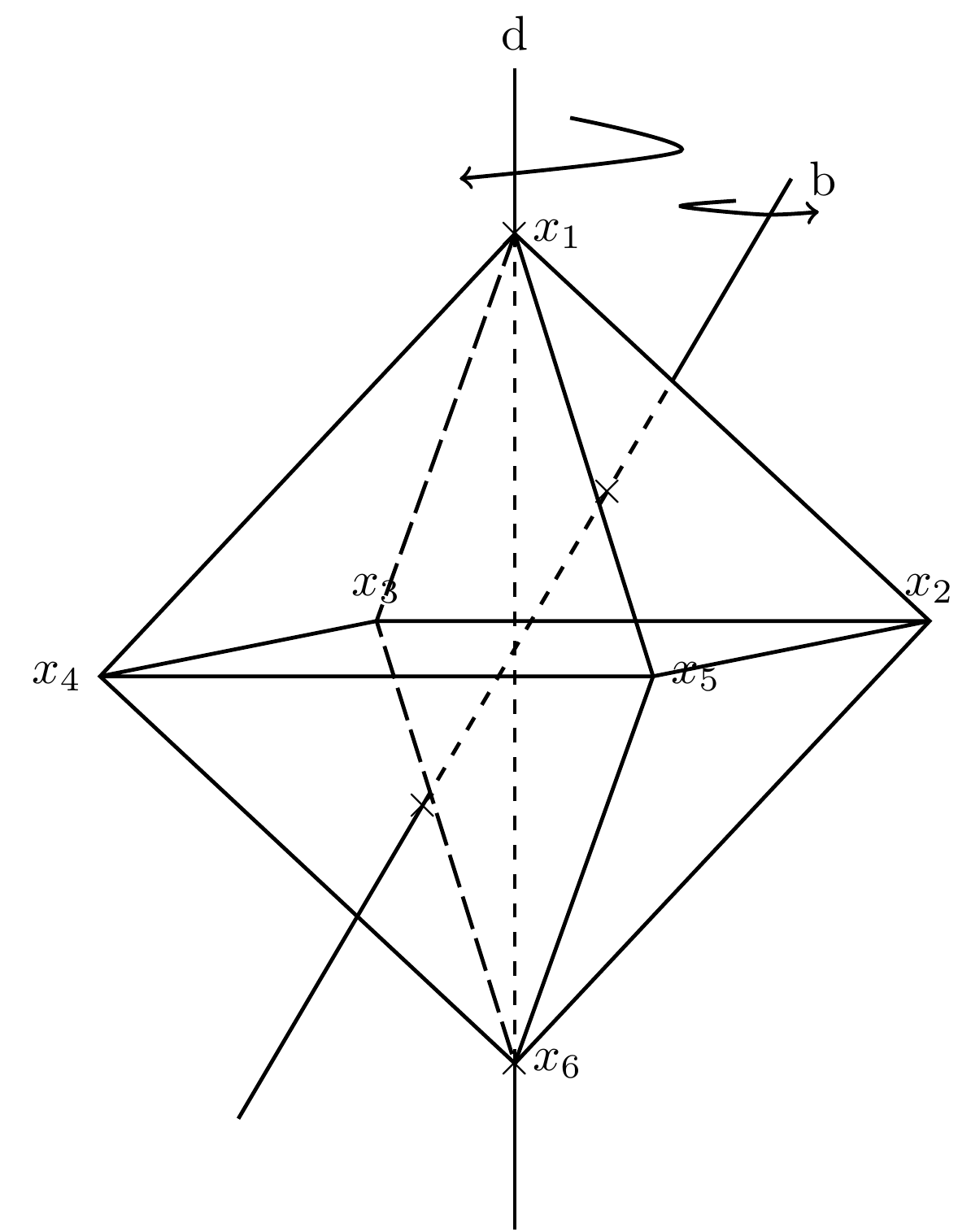}
    \caption{Octahedron and Cube are dual to each other. We parameterize Octahedron similar to the cube,  $b^3=d^4=e$, $db^2d=b$ and $dbd=bd^2b$. The six vertices exchange under $S_4$ as $b: x_1 \mapsto x_2 \mapsto x_3 \mapsto x_1, x_4\mapsto x_5\mapsto x_6 \mapsto x_4$, $d: x_1\mapsto x_1, x_6\mapsto x_6, x_3\mapsto x_2 \mapsto x_5\mapsto x_4 \mapsto x_3 $.}
    \label{fig: O}
\end{figure}

The six dimensional exchange representation can be decomposed as 
\begin{equation}
    \begin{split}
        \mathbf{6}&=\mathbf{1}\oplus\mathbf{2}\oplus\mathbf{3'}\\
         \begin{pmatrix}
    x_1\\ x_2 \\ x_3\\ x_4 \\x_5\\x_6 \end{pmatrix}_\mathbf{6}^\text{exchange}=&\begin{pmatrix} x_1+x_2+ x_3+ x_4+x_5+x_6\end{pmatrix}_\mathbf{1}\\
    & \oplus\begin{pmatrix}(x_1+ x_6)+ \omega (x_2+x_4)+ \omega^2(x_3+ x_5) \\ (x_1+ x_6)+ \omega^2 (x_2+x_4) + \omega(x_3+ x_5) \end{pmatrix}_\mathbf{2}\\
    &\oplus\begin{pmatrix} \sqrt{3}(x_1-x_6)\\\sqrt{3}(x_2-x_4)  \\\sqrt{3}(x_3-x_5) \end{pmatrix}_\mathbf{3'}
    \end{split}
\end{equation}
where $\omega=\exp(\frac{2i\pi}{3})$.
\subsection{Icosahedron}
Icosehedron~\cite{Ishimori:2010au} and Dodecahedron are dual to each other. Icosehedron has 12 vertices where as Dodecahedron has 20. 
For simplicity, we will only consider the Icosehedron.
%We will only present the reduction of fields represented as the vertices as the Icosehedron into irreducible representation.
\begin{figure}[H]
    \centering
    \includegraphics[width=0.55\textwidth]{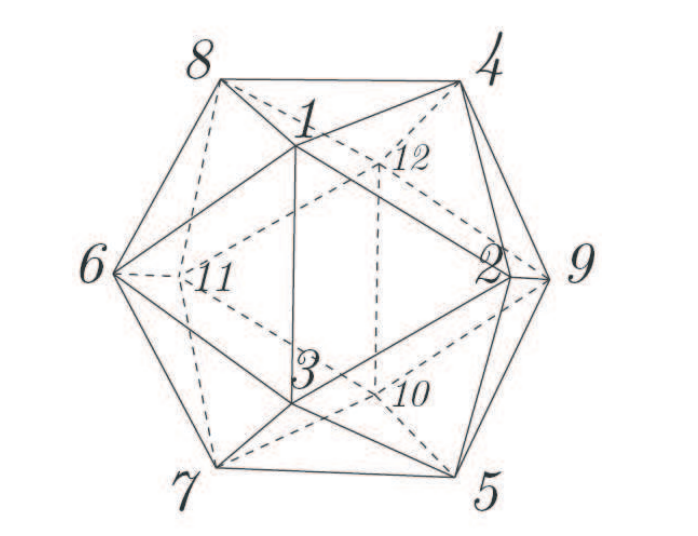}
    \caption{Proper rotations of an Icosehedron is isomorphic to $A_5$. $A_5$ can be parameterized by~\cite{Ishimori:2010au} $a^2=b^5=(ab)^5=e$, where $a$ corresponds to rotation by $\pi$ about the edge joining vertices 1 and 2 and $b$ corresponds to clockwise rotation by axis passing through the centre of the face 10-11-12. The 12 vertices exchange under $A_5$ as $a: (x_1,x_2,x_3,x_4,x_5,x_6,x_7,x_8,x_9,x_{10},x_{11},x_{12})\mapsto(x_2,x_1,x_4,x_3,x_8,x_9,x_{12},x_5,x_6,x_{11},x_{10},x_7)$, $b: (x_1,x_2,x_3,x_4,x_5,x_6,x_7,x_8,x_9,x_{10},x_{11},x_{12})\mapsto(x_2,x_3,x_1,x_5,x_6,x_4,x_8,x_9,x_7,x_{11},x_{12},x_{10})$.}
    \label{fig:Icosehedron}
\end{figure}

This 12 dimensional exchange representation can be decomposed as 
\begin{equation}
    \begin{split}
        \mathbf{12}&=\mathbf{1}\oplus\mathbf{3}\oplus\mathbf{3'}\oplus\mathbf{5}\\
         \begin{pmatrix}
    x_1\\ x_2 \\ x_3\\ x_4 \\x_5\\x_6\\x_7\\x_8\\x_9\\x_{10}\\x_{11}\\x_{12} \end{pmatrix}_\mathbf{12}^\text{exchange}=&\begin{pmatrix} x_1+x_2+ x_3+ x_4+x_5+x_6+x_7+x_8+x_9+x_{10}+x_{11}+x_{12}\end{pmatrix}_\mathbf{1}\\
    & \oplus\begin{pmatrix}x_2+x_5-x_8-x_{11}+\frac{1}{\phi}(x_3+x_9-x_6-x_{12})\\ x_1+ x_4- x_7- x_{10}+\frac{1}{\phi}(x_2+x_8-x_5-x_{11})\\ x_3+ x_6- x_9- x_{12}+\frac{1}{\phi}(x_1+x_7-x_4-x_{10})\end{pmatrix}_\mathbf{3}\\
    &\oplus\begin{pmatrix}x_1+ x_4- x_7- x_{10}+\phi(x_5+x_{11}-x_2-x_8)\\x_8+x_{11}-x_2-x_5+\phi(x_3+x_9-x_6-x_{12})  \\ x_3+ x_6- x_9- x_{12}+\phi(x_4+x_{10}-x_1-x_7) \end{pmatrix}_\mathbf{3'}\\
    &\oplus\begin{pmatrix}x_1+x_4+x_7+x_{10}-\frac{1}{\phi}(x_2+x_5+x_8+x_{11})+(\frac{1}{\phi}-1)(x_3+x_6+x_9+x_{12})\\\frac{1}{\phi}(x_2+x_{11}-x_5-x_8) \\\frac{1}{\phi}(x_1+x_{10}-x_4-x_7) \\ \frac{1}{\phi}(x_3+x_{12}-x_6-x_9)\\a_1(x_1+x_4+x_7+x_{10})+a_2(x_2+x_5+x_8+x_{11})+a_3(x_3+x_6+x_9+x_{12}) \end{pmatrix}_\mathbf{5}
    \end{split}
\end{equation}
 where $\phi=\frac{\sqrt{5}+1}{2}$, $a_1=\frac{3-2\phi}{\sqrt{3}}$, $a_2=\sqrt{\frac{3}{5}}\frac{1+2\phi}{\phi^4}$ and $a_3=\frac{1+5\phi^3}{2\sqrt{3}(\phi+2\phi^2)}$.
\end{appendices}

\bibliographystyle{unsrtnat}
\bibliography{mybibliography.bib}

%\end{multicols}
\end{document}